\documentclass[a4paper,11pt]{article}
\usepackage{jcappub} % for details on the use of the package, please
\usepackage[T1]{fontenc} % if needed

\usepackage{graphicx}
\usepackage{slashed}
\usepackage{soul}
\usepackage{ulem}
\usepackage{color}

\newcommand{\eq}[1]{Eq.~(\ref{#1})}

\newcommand{\MeV}{\mathinner{\mathrm{MeV}}}
\newcommand{\GeV}{\mathinner{\mathrm{GeV}}}
\newcommand{\TeV}{\mathinner{\mathrm{TeV}}}
\newcommand{\PeV}{\mathinner{\mathrm{PeV}}}

\def\l{\left}
\def\r{\right}
\def\pl{{\rm P}}

\newcommand{\beq}{\begin{equation}}
\newcommand{\eeq}{\end{equation}}
\newcommand{\bea}{\begin{eqnarray}}
\newcommand{\eea}{\end{eqnarray}}

\title{\boldmath Hidden sector monopole, vector dark matter and dark radiation with Higgs portal}

\author{Seungwon Baek,}
\author{P. Ko}
\author{and Wan-Il Park}
\affiliation{School of Physics, KIAS,\\Seoul 130-722, Korea}

\emailAdd{sbaek1560@gmail.com}
\emailAdd{pko@kias.re.kr}
\emailAdd{wipark@kias.re.kr}

\abstract{We show that the 't Hooft-Polyakov monopole model in the hidden sector 
with Higgs portal interaction makes a viable dark matter model, where monopole and 
massive vector dark matter (VDM) are stable due to topological conservation and the unbroken   subgroup $U(1)_X$.  
We show that, even though observed CMB data requires the dark gauge coupling to be 
quite small, a right amount of VDM thermal relic can be obtained via $s$-channel resonant annihilation for the mass of VDM close to or smaller than the half of SM higgs mass, 
thanks to Higgs portal interaction.
Monopole relic density turns out to be several orders of magnitude smaller than observed 
dark matter relic density.
Direct detection experiments, particularly, the projected XENON1T experiment, may probe 
the parameter space where the dark Higgs is lighter than $\lesssim 60 \GeV$. 
In addition, the dark photon associated with unbroken $U(1)_X$ contributes to the radiation energy density at present, giving $\Delta N_{\rm eff}^\nu \sim 0.1$ as the extra relativistic 
neutrino species.}

\begin{document}
\maketitle
\flushbottom

\section{Introduction} 
One of the mysteries of the universe is that 26 \% of it is made of nonbaryonic cold 
dark matter (CDM) \cite{Ade:2013zuv}, which definitely calls for beyond the SM (BSM)  
physics. 
As of now, very little is known about the particle nature of CDM except that it carries 
no electric or color charge. We do not know (i) how many species of CDM's are there, 
(ii) if DM is absolutely stable or have very long lifetime, (iii) what are the masses and 
spins of CDM's, and (iv) how CDMs interact with each other or the ordinary matter.  
This lack of information results in many models for dark matter. Very often some ad hoc 
$Z_2$ symmetry or similar is introduced in order to stabilize DM without questioning the  
origin of those symmetries.  
Also, DM particles often feel no gauge interaction, unlike most of the SM particles. 
However weakly interacting massive particle (WIMP) with mass $\sim O(100)$ GeV, 
which is  the most common CDM candidate, is not likely to be stable under a global 
symmetry which is often assumed for the stability of DM \cite{singlet_portal}.  
The stability of DM and the fact that SM is guided by local gauge principle may imply that 
the dark sector in which the dark matter responsible for the present relic density resides 
may respect local gauge symmetry, too. This picture also arises naturally in 
string inspired models \cite{Shiu:2013wxa}.

The last unbroken dark gauge symmetry ($H_X$) guaranteeing the stability of DM 
may be originated from a larger gauge group ($G_X$).
In this case, topologically stable objects are likely to form during the symmetry-breaking 
phase transition, although it depends on the nature 
of the larger gauge group and the pattern of the symmetry breaking.  
For example one can have topological monopole if $\Pi_2 ( G_X / H_X ) = Z$ (integer), vortices (strings) or domain walls depending on the lower homotopy classes. 
Moreover, since the hidden sector may communicate with the visible sector via various gauge singlet 
portal interactions~ \cite{singlet_portal}, even topological soliton(s) may have a chance to leave observable imprints.

In this paper, we consider a simple hidden sector DM model, where non-Abelian dark gauge symmetry $SU(2)_X$ is broken down to a $U(1)_X$ by a real triplet dark Higgs field.
It is just the 't Hooft-Polyakov monopole \cite{'tHooft:1974qc,Polyakov:1974ek} model 
in the hidden sector.  In this well known setup, we add the Higgs portal interaction 
which is allowed at renormalizable level, and show that a viable dark matter phenomenology can be obtained. 

This paper is organized as follows.
In section~\ref{sec:model}, our model is proposed and particle spectra in the model is discussed.
In section~\ref{sec:low-pheno}, we discuss about constraints coming from the vacuum stability, perturbative unitarity and collider data.
Section~\ref{sec:DM-pheno} is devoted to DM-phenomenology, where constraints on DM self-interaction from formation of massive black holes and CMB data are discussed, and relic densities of VDM and monopoles are estimated.
In section~\ref{sec:DR}, dark radiation is discussed briefly, and conclusion is drawn in section~\ref{sec:conc}.

%---------------------------------------------------------------------------------------------------------------------
\section{Model and Particle Spectra} \label{sec:model}
%---------------------------------------------------------------------------------------------------------------------
Let us consider $SU(2)_X$-triplet real scalar field $\vec{\Phi}$ with the following Lagrangian:
\begin{equation} \label{Lag}
{\cal L} = {\cal L}_{\rm SM} - \frac{1}{4} V_{\mu\nu}^a V^{a \mu\nu} + \frac{1}{2} 
D_\mu \vec{\Phi} \cdot D^\mu \vec{\Phi}
-  \frac{\lambda_\Phi}{4} 
\left( \vec{\Phi} \cdot \vec{\Phi} - v_\phi^2 \right)^2 
- \frac{\lambda_{\Phi H}}{2}   \left(\vec{\Phi} \cdot \vec{\Phi} - v_\phi^2\right)
\left( H^\dagger H - \frac{v_H^2}{2} \right) 
\end{equation}
where ${\cal L}_{\rm SM}$ is the standard model Lagrangian, $D_\mu \Phi^a = \partial_\mu \Phi^a - g_X  \epsilon^{abc} V_\mu^b \Phi^c$ and $V_{\mu\nu}^a = \partial_\mu V_\nu^a - \partial_\nu V_\mu^a - g_X  \epsilon^{abc} V_\mu^b V_\nu^c$ with $\epsilon^{abc}
(a,b,c=1,2,3)$ being the structure constant of the hidden sector $SU(2)$ gauge group.
The Higgs portal interaction is described by the $\lambda_{\Phi H}$ term. 
When we ignore the Higgs portal interaction, the hidden sector Lagrangian describes 
the 't Hooft-Polyakov monopole~\cite{'tHooft:1974qc,Polyakov:1974ek}.
After the spontaneous symmetry breaking of $SU(2)_X \approx SO(3)_X$ into 
$ U(1)_X \approx SO(2)_X $ 
by nonzero vacuum expectation value (VEV) of $\vec{\Phi}$, 
\[
\langle \vec{\Phi} (x) \rangle = ( 0 , 0 , v_\Phi ) , 
\] 
hidden sector particles are composed of massive dark vector bosons $ V_\mu^\pm \equiv \l( V_\mu^1 \mp i V_\mu^2 \r)/\sqrt{2}$ with masses $m_V=g_X v_\Phi$ 
\footnote{Here $\pm 1$ indicate the dark charge under $U(1)_X$, and not ordinary electric 
charges.}, massless dark photon $\gamma_{h, \mu} \equiv V_\mu^3$, massive real scalar 
$\phi$ (dark Higgs boson) and topologically stable heavy (anti-)monopole  with mass 
$m_M=m_V/\alpha_X$.  

After the spontaneous breaking of electroweak symmetry, Higgs portal interaction mixes 
$\phi$ and SM Higgs boson $h$.    After imposing the vanishing tadpole conditions, 
the mass$^2$ mixing between $h$ and $\phi$  is described by the following matrix: 
\begin{equation}
\left( \begin{array}{cc} 
        m_{hh}^2 & m_{\phi h}^2 \\
        m_{\phi h}^2 & m_{\phi \phi}^2 \\
        \end{array}   \right)
\equiv
\left( \begin{array}{cc} 
        2 \lambda_H v_H^2 & \lambda_{\phi H} v_H v_\phi \\
        \lambda_{\phi H} v_H v_\phi & 2 \lambda_\phi v_\phi^2 \\
        \end{array}   \right)
\end{equation}
in the $( h, \phi)$ basis with $\lambda_H$ being the quartic coupling of the SM Higgs.  
We can make a $SO(2)$ rotation from $(h , \phi)$ basis 
to the physical mass eigenstates, $( H_1 , H_2 )$ with  mass eigenvalues
\beq
m_{1,2}^2 = \frac{1}{2} \l[ \l( m_{hh}^2 + m_{\phi \phi}^2 \r) \mp \sqrt{\l( m_{hh}^2 - m_{\phi \phi}^2 \r)^2 + 4 m_{\phi h}^4 } \r]
\eeq
Note that $m_{1,2}^2 > 0$ requires
%\beq \label{stability-cond}
$\lambda_{\phi H} < 2 \sqrt{\lambda_H \lambda_\phi}$.
%\eeq
With the mixing angle $\alpha$ defined by
\beq \label{mixing-angle}
\tan 2 \alpha = \frac{2 m_{\phi H}^2}{m_{hh}^2 - m_{\phi\phi}^2}
\eeq
the interaction eigenstates can be expressed in terms of mass eigenstates as
\begin{equation}
\left( \begin{array}{c} h \\ \phi \end{array} \right)
=
\left( \begin{array}{cc} \cos \alpha & -\sin \alpha \\ \sin \alpha & \cos \alpha \end{array} \right)
\left( \begin{array}{c} H_1 \\ H_2 \end{array} \right)
\end{equation}
This is similar to the renormalizable models for singlet fermion DM \cite{fermion_dm,vacuum_stability} or VDM~\cite{vector_dm}.  

Note that there is no kinetic mixing between $\gamma_h$ and the SM 
{$U(1)_Y$-gauge boson unlike  the $U(1)_X$-only case,   
due to the non Abelian nature of the hidden gauge symmetry. 
The hidden vector bosons $V_\mu^\pm$ are absolutely stable due to the unbroken 
$U(1)_X$ gauge symmetry even if we consider nonrenormalizable interactions.   Hence they 
become good CDMs in addition to monopoles without additional dark charged matter fields. 
This aspect is in sharp contrast with the VDM with the $SU(2)_X$ being completely broken~\cite{hambye}, where the stability of massive VDM is not protected by 
$SU(2)_X$ gauge symmetry and nonrenormalizable interactions would make the VDM 
decay in general. 
In the model presented in this paper, the unbroken $U(1)_X$ subgroup not only protects the stability of VDM $V_\mu^\pm$, but also contributes to the dark radiation at the level of 
$\sim 0.1$.

%---------------------------------------------------------------------------------------------------------------------
\section{Low energy phenomenology} \label{sec:low-pheno}
The presence of symmetry breaking \textit{dark higgs} field and \textit{Higgs portal} 
interaction allows a mixing between SM- and dark-Higgs bosons.
This mixing improves the vacuum stability of SM Higgs potential, but is constrained 
by collider experiments as described below.

\subsection{Vacuum Stability and Perturbative Unitarity} 
The Higgs portal interaction of $\Phi$ to the SM Higgs $H$ can improve the vacuum 
stability along SM Higgs direction up to Planck scale via tree-level $h-\phi$ mixing 
and additional scalar loop-correction to the quartic coupling of SM Higgs.
As shown in Ref.~\cite{vacuum_stability}, for $m_t = 173.2 \GeV$ and $\alpha_s = 0.118$ 
as the top-quark pole mass and strong gauge coupling, the vacuum instability 
can be cured  up to Planck scale, if  $m_1 < m_2$ and the scalar mixing angle ($\alpha$) satisfies 
\beq \label{lH}
\lambda_H =   \lambda_H^{\rm SM}\l[ 1 - \l( 1 - \frac{m_2^2}{m_1^2} \r) \sin^2(\alpha) \r] \gtrsim 0.139  ,
\eeq
where $\lambda_H^{\rm SM} \equiv m_h^2 / \l( 2 v_H^2 \r) \simeq 0.129$ is the Higgs quartic coupling of SM with $m_1 = m_h \approx 125 \GeV$ being the mass of the observed SM highs-like particle, or $0.2 \lesssim \lambda_{\Phi H} \lesssim 0.6$ with $0 \leq \lambda_\Phi \lesssim 0.2$.

Recent result from LHC experiments constrains the mixing angle to be
$\alpha \lesssim 0.45$ at $95$\% CL~\cite{Chpoi:2013wga}.
For such a small mixing, the mass eigenstates can be approximated to the interaction eigenstates, and in case of absolute stability \eq{lH} is translated to $m_\phi \gtrsim 150 \GeV$ if $\lambda_{\Phi H}$ is too small to improve the vacuum stability.
Noth that, if $m_2 < m_1$, vacuum instability becomes worse for a non-zero $\alpha$, and hence somewhat large $\lambda_{\Phi H}$ would be necessary.
However, for a small $\alpha$ satisfying $\tan \alpha \lesssim m_\phi / m_h$, tachyon-free condition $\lambda_{\Phi H} < 2 \sqrt{\lambda_\Phi \lambda_H}$ is translated to 
\beq \label{lphiHbnd}
\lambda_{\Phi H} \lesssim \sqrt{\frac{g_X m_\phi}{m_V} \l[ 1 - \l( 1 - \frac{m_\phi^2}{m_h^2} \r) \sin^2 \alpha \r] \frac{m_h^2}{v_H^2}}
\eeq
implying that the small $m_V$ is, the larger $\lambda_{\Phi H}$ can be.
In addition, as will be discussed in section~\ref{subsec:CMB-const} and~\ref{sec:relics}, in order to be consistent with CMB constraint and to obtain a right amount of relic density, one needs 
\beq
2 m_V \approx m_\phi \ {\rm or} \ m_h
\eeq
Moreover, for such a light VDM, $g_X$ is constrained by small scale structure as \eq{alphaX-bound}.
Hence, for $m_\phi < m_h$ and $2 m_V \approx m_\phi$, \eq{lphiHbnd} can be written as 
\beq \label{lphiHbnd2}
\lambda_{\Phi H} \lesssim 7.6 \times 10^{-2} \l( \frac{m_V}{50 \GeV} \r)^{3/8} \l[ 1 - \l( 1 - \frac{4 m_V^2}{m_h^2} \r) \sin^2 \alpha \r]^{1/2}
\eeq
which looks a bit small to cure vacuum instability problem.
Therefore, in case of $m_\phi < m_h$, SM vacuum might be meta-stable modulo uncertainties in the pole mass of top quark and the strong coupling constant $\alpha_s$.

%-------------------------------------------------------------------------------------------------------------
\subsection{Phenomenology of Two Scalar Bosons}
%As shown in Fig.~\ref{fig:sp-vs-mphi}, a wide range of $m_\phi$ can match present observations and {\color{red} some region} may be probed in the near future.
%Particulraly, 
For $m_\phi > 2 m_h$, the decay channel of 4-charged-lepton final states is open (see \cite{vector_dm} for details) and provide a clean signal.
It might be within the reach of LHC experiments. 
On the other hand, if $m_\phi < m_h/2$, SM higgs can decay into two lighter 
dark Higgs bosons which decay subsequently to light SM particles.
Since $\lambda_\phi$ and $\lambda_{\phi H}$ is much smaller than $\lambda_H$ in this case (see the next section), the decay of SM Higgs to dark Higgs is mainly due to $\lambda_H$ coupling, 
and the decay rate is found to be
\beq
\Gamma_{h \to \phi \phi} = \frac{9 \l( c_\alpha s_\alpha^2 \r)^2}{32 \pi} \frac{m_h^3}{v_H^2} \l( 1 - \frac{4 m_\phi^2}{m_h^2} \r)^{1/2}
\eeq
where $m_h \simeq \sqrt{2 \lambda_H} v_H$ was used.
The subsequent decay rate of $\phi$ would be the same as that of SM-like Higgs boson,  
with the replacement $m_h \leftrightarrow m_\phi$ and a universal suppression factor 
$\sin^2 \alpha$ in the decay rate for each channel. 
Therefore, the decay of SM-like Higgs to dark Higgs bosons will produce dominantly 
4 b-jets,  if kinematically allowed ($m_\phi > 2 m_b \sim 10$ GeV).
LEP Higgs search imposes a bound on such a process and requires $\alpha \lesssim 0.3$ 
for $m_\phi \lesssim 60 \GeV$, as shown in Fig.~\ref{fig:LEP-bnd}.
%---------------------
\begin{figure}
\centering
\includegraphics[width=0.45\textwidth]{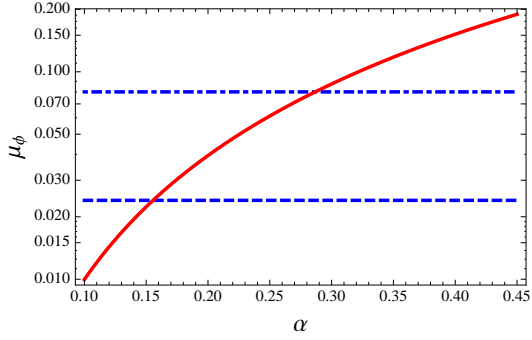}
\caption{\label{fig:LEP-bnd} 
Signal strength (solid red line) of SM channels caused by the production and decay of dark higgs, as a function of $\alpha$.
Blue dashed and dot-dashed lines are the upper-bound at 95\% CL for $m_\phi \approx 20, 60 \GeV$, respectively \cite{Barate:2003sz}.
}
\end{figure}
%---------------------
LHC experiments also impose a bound on the branching fraction of the SM-like Higgs 
boson into non-SM channels, as shown in Fig.~\ref{fig:LHC-bnd} where we used an approximation \cite{Chpoi:2013wga}
\[
c_\alpha \geq 0.904 + {\rm Br}_{\rm non-SM} /2
\] 
with ${\rm Br}_{\rm non-SM}$ being the branching fraction of the decay of SM-like higgs to non-SM channels.
%Fig.~\ref{fig:BrHtoXX-vs-alpha} shows branching fractions of SM-like higgs to VDM(red), dark higgs(blue) and sum of both channels, as functions of $\alpha$.
%In the upper-panel of the figure, green lines are nearly overlapped with blue lines because of the relative smallness of VDM channel.
%---------------------
\begin{figure}
\centering
\includegraphics[width=0.45\textwidth]{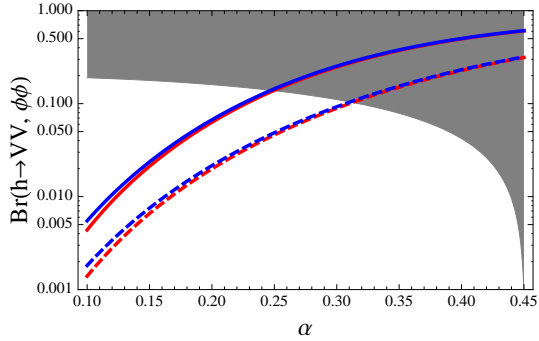}
\caption{\label{fig:LHC-bnd} 
Branching fraction of the decay of SM-like Higgs to non-SM (two VDMs or dark higgs), as a function of $\alpha$.
Red: $m_V = \l(1- 3/40 \r) \times m_h/2$, Blue: $m_V = \l(1- 3/40 \r) \times m_\phi/2$.
To be consistent with the constraint described in the next section, $g_X = 4.5 \times 10^{-2} \l( 2 m_V/ 1 \TeV \r)^{3/4}$ was used. Solid and dashed lines correspond to $m_\phi = 20, 60 \GeV$, respectively.
Gray region is excluded by collider experiments at 95\% CL \cite{Chpoi:2013wga}.}
\end{figure}
%---------------------

Note that for nonzero scalar mixing angle ($\alpha \neq 0$), the signal strength of the SM Higgs-like scalar boson is 
less than 1 in a universal manner~\cite{singlet_portal,fermion_dm,vector_dm}, 
which is a generic in Higgs portal DM models with only one Higgs doublet as in the SM.

%-------------------------------------------------------------------------------------------------------------
\section{Dark matter phenomenology} \label{sec:DM-pheno}

The massless dark photon ($\gamma_h$) in our scenario mediates a long range force 
between vector dark matters ($V_\mu^\pm$) and (anti)monopoles.
Particularly, such a massless mediator can cause a large non-perturbative enhancement of perturbative pair annihilation or self-interactions of dark matter.
The enhancement, named as Sommerfeld enhancement, is given by \cite{Sommerfeld:1931,ArkaniHamed:2008qn}
\beq
\mathcal{S} = \frac{\pi \alpha_X / v}{1 - e^{-\pi \alpha_X / v}}
\eeq
where $v$ is the velocity of dark matter.
Note that, when $v \ll \alpha_X$, the enhancement is proportional to $1/v$.
This behavior of $\mathcal{S}$ has crucial impacts on CMB and physics of small scale dark matter subhalos, as described in the following two subsections.
For simplicity, we begin with the constraint from small scale dark matter subhalos.

\subsection{Self interaction between dark matters}
The dark photon carries a long range dark force between monopoles and VDMs.
For the main component of dark matter, the self-interaction is strongly constrained by small and large scale structures (dwarf galaxies, bullet cluster, etc.), similarly to the singlet portal scalar DM considered in Ref.s~\cite{singlet_portal,Feng:2009mn}.
Since monopole contribution to dark matter of the universe  turns out to be subdominant 
as described in the section~\ref{sec:relics}, we will consider only the case of VDM 
self-interactions in this subsection.

The transfer cross section of the VDM self-interaction mediated by dark photon is 
\beq \label{sT}
\sigma_T = \mathcal{S} \times \frac{\pi \alpha_X^2}{m_V^2 v_{\rm cm}^4} \ln \l[ \frac{m_V^2 v_{\rm cm}^3}{\l( 4 \pi \rho_V \alpha_X^3 \r)^{1/2}} \r]
\eeq
where $v_{\rm cm}$ and $\rho_V$ are respectively the velocity and energy density of dark matter at the region of interest, and 
only the contribution from attractive interaction was included since repulsive interaction causes suppression rather than enhancement.
Formation of massive blackholes caused by DM self-interaction \cite{Ostriker:1999ee} may impose the most strong bound  \cite{Loeb:2010gj}:
\beq \label{sTbnd}
\frac{\sigma_T}{m_V} \lesssim \l( \frac{\sigma_T}{m_V} \r)^{\rm max} \equiv 35 {\rm cm}^2/{\rm g} \ {\rm  for} \ v_{\rm cm} = 10 {\rm km/s}
\eeq 
leading to an upper-bound on $\alpha_X$ that is depicted in Fig.~\ref{fig:alphaX-bnd} as a function of $m_V$.
%\footnote{In fact, in our scenario, the gravothermal catastrophe is not likely to occur since the transfer cross section of VDM self-interaction at a relativistic regime is highly suppressed due to $v_{\rm cm}^{-4}$ dependence of $\sigma_T$.
%This constraint may be necessary to avoid too large core of dwarf galaxy scale dark matter subhalos.
%}.
%---------------------
\begin{figure}
\centering
\includegraphics[width=0.45\textwidth]{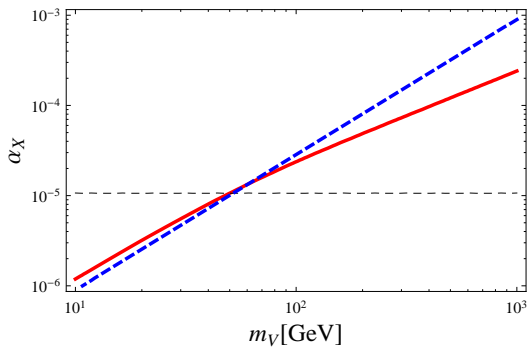}
\caption{\label{fig:alphaX-bnd} 
Bound on $\alpha_X$ (red line) obtained from \eq{sTbnd} as a function of $m_V$.
Dashed blue line is the case when enhancement factor is ignored.
Horizontal dashed black line corresponds to $\pi \alpha_X / v_{\rm cm}=1$ with $v_{\rm cm}=10 {\rm km}/{\rm sec}$ being the velocity of dark matter at subhalos of dwarf galaxy scale.}
\end{figure}
%---------------------
For $\alpha_X \lesssim v/\pi$ leading to $\mathcal{S} \sim 1$, the constraint can be interpreted approximately as 
\beq \label{alphaX-bound}
%g_X \lesssim g_X^{\rm max}(m_V) \equiv 1.7 \times 10^{-2} \l( \frac{m_V}{100 \GeV} \r)^{3/4}
\alpha_X \lesssim 10^{-5} \l( \frac{m_V}{50 \GeV} \r)^{3/2}
\eeq
where $v_{\rm cm} = 10 {\rm km/s}$ and $\rho_V = 3 \GeV/{\rm cm}^3$ were used.
In fact, \eq{alphaX-bound} is valid for $10 \GeV \lesssim m_V \lesssim 100 \GeV$, which will be the range of our prime interest, within $\mathcal{O}(10)$\% error.
%
%
%{\color{red} 
%In the case of monopoles, for the magnetic charge of monopole $g_{\rm M} \equiv 4 \pi/g_X$, if interactions between monopoles are still perturbative, i.e., $g_{\rm M} \lesssim \sqrt{4 \pi}$,  we can apply the same argument to the self-interaction of monopoles with replacements, $g_X \leftrightarrow g_{\rm M}$ and $m_V \leftrightarrow m_{\rm M}$.
%In this case, we obtain a new constraint
%\beq
%\sqrt{4 \pi} \lesssim g_X \lesssim g_X^{\rm max}(m_V)^2/\sqrt{4 \pi}
%\eeq
%which has an allowed space only for $m_V \gtrsim 116 \TeV$.
%
%}
%
%
%{\color{red} For $g_X \ll 1$, $g_{\rm M} \gg \mathcal{O}(10)$ and causes a very strong self-interaction of monopoles.
%In this case, monopole contribution to dark matter may have to be small enough in order to be consistent with observed structures at small and large scales.}
%
%{\color{red} The abundance of such strongly interacting monopoles may be constrained by CMB and large scale structure \cite{Cyr-Racine:2013fsa}, and/or possible effects of self-annihilation caused by capture at sun \cite{Zentner:2009is}.
%Since the detailed analysis is out of the scope of this paper, here we simply assume that the abundance of monopoles should be less than 5\% of the observed CDM relic density, based on the result of Ref. \cite{Cyr-Racine:2013fsa}.}
%

Interestingly, for $\alpha_X^{\rm max} /10^2 \lesssim \alpha_X \lesssim \alpha_X^{\rm max}$ with $\alpha_X^{\rm max}$ being the uppper-bound of $\alpha_X$, self-interactions among VDM can resolve the core/cusp problem and ``too-big-to-fail'' problem of the conventional collisionless CDM scenarios \cite{Loeb:2010gj,Vogelsberger:2012ku}.

\subsection{Constraint from CMB} \label{subsec:CMB-const}
Around the epoch of CMB decoupling, the velocity of dark matter is given by
\beq
v_{\rm cmb} = v' \l[ \frac{g_{*S}(T_{\rm cmb})}{g_{*S}(T')} \r]^{1/3} \frac{T_{\rm cmb}}{T'}
\eeq
where $v$ is the velocity of dark matter, $T$ is the photon temperature, and `$'$' and the subscript `$_{\rm cmb}$' denote the epochs of DM's last kinetic decoupling and CMB decoupling, respectively.
If DM is kinetically decoupled from visible sector at $T = T_{\rm kd}$ while it is still coupled to dark photon, the thermal bath around the epoch of the last kinetic decoupling is provided by dark photon.
In this case, the temperture of dark photon $T_{\gamma'}$ for $T<T_{\rm kd}$ is given by
\beq \label{Tg-to-Tgh}
T_{\gamma'} = \l( \frac{g_{*S}(T)}{g_{*S}(T_{\rm kd})} \r)^{1/3} T
\eeq
The Compton scattering rate of DM to dark photon when the temperature of dark photon is $T_{\gamma'}$ is 
\beq
\Gamma_{\rm Comp} = \frac{32 \pi^3 \alpha_X^2 T^4}{45 m_V^3} \l( \frac{g_{*S}(T)}{g_{*S}(T_{\rm kd})} \r)^{4/3}
\eeq
Comparing to the expansion rate, $H = \l( \pi^2 g_*(T)/90 \r)^{1/2} T^2/M_\pl$, one finds that the photon temperature when DM is kinetically decoupled from dark photon is
\beq \label{Tprime-kd}
T'
= \l( \frac{45}{32 \pi^3 \alpha_X^2} \r)^{1/2} \l( \frac{g_{*S}(T')}{g_{*S}(T_{\rm kd})} \r)^{-2/3} \l( \frac{\pi^2}{90} g_*(T') \r)^{1/4} \l( \frac{m_V}{M_\pl} \r)^{3/2} M_\pl
\eeq
Hence, using $v' = \sqrt{3 T_{\gamma'}'/m_V}$ and \eq{Tg-to-Tgh}, one finds
\beq
v_{\rm cmb}
= \l( \frac{32 \pi^3 \alpha_X^2}{5} \r)^{1/4} \l( \frac{3 \sqrt{10}}{\pi} \r)^{1/4} \frac{g_{*S}(T')^{1/24} g_{*S}(T_{\rm cmb})^{1/3} }{g_{*S}(T_{\rm kd})^{1/2}} \l( \frac{T_{\rm cmb}}{M_\pl} \r) \l( \frac{M_\pl}{m_V} \r)^{5/4}
\eeq
where we used $g_*(T') = g_{*S}(T')$.
Fig.~\ref{fig:Tkd-vdm} shows temperatures of photon and dark photon (left panel), and $v_{\rm cmb}$ (right panel) at the last kinetic decoupling as functions of $m_V$.
%---------------------
\begin{figure}
\centering
\includegraphics[width=0.45\textwidth]{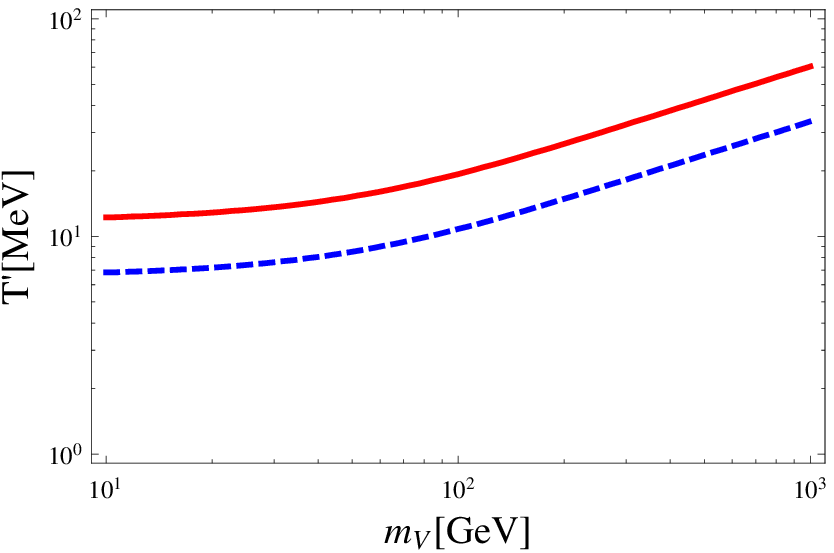}
\includegraphics[width=0.45\textwidth]{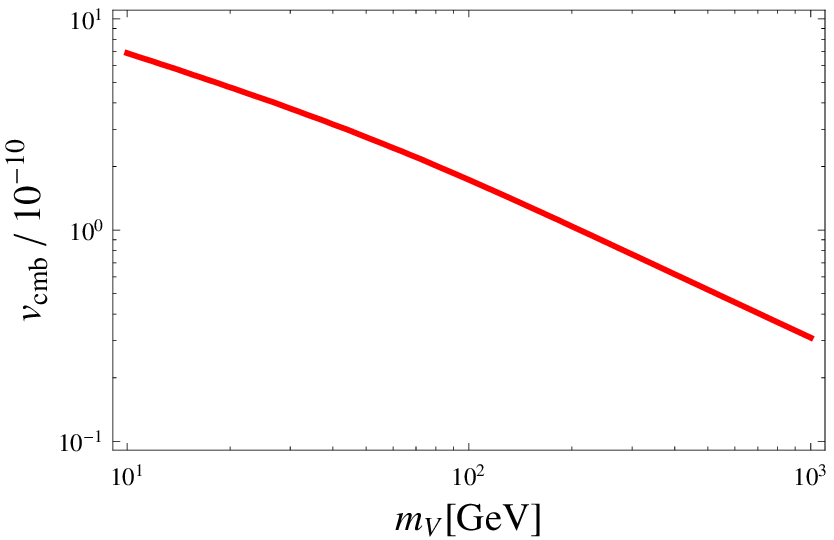}
\caption{\label{fig:Tkd-vdm} 
Left: Temperatures of photon ($T'$: red solid line) and dark photon ($T_{\gamma'}'$: blue dashed line) at the last kinetic decoupling of dark matter.
Right: Velocity of dark mater around the epoch of CMB decoupling ($v_{\rm cmb}$). 
We used $\alpha_X$ saturating the bound from small scale structures.}
\end{figure}
%---------------------

The present CMB data constrains the velocity-averaged annihilation cross section of dark matter to SM particles to be upper-bounded at \cite{Madhavacheril:2013cna}
\bea \label{CMB-const}
\langle \sigma v_{\rm rel} \rangle_{\rm tot}^{\rm cmb} 
&=& \frac{0.66 \times 10^{-6} \ m_V}{\sum_{i={\rm channels}} f_{\rm eff, sys}^i {\rm Br}^i} \l( \frac{{\rm m}^3}{\rm sec \cdot kg} \r)
\nonumber \\
&\simeq& \frac{1.2 \times 10^{-24}}{\sum_{i={\rm channels}} f_{\rm eff, sys}^i {\rm Br}^i} \l( \frac{m_V}{1 \TeV} \r) {\rm cm}^3/{\rm sec}
\eea
where $f_{\rm eff,sys}^i$ and ${\rm Br}^i$ are respectively the fractional energy deposition of annihilation products and the branching faction of channel $i$.
%As an example, for $m_V \approx m_h/2$, we may use
%\beq
%f^{bb} = 0.23, {\rm Br}^{bb} \simeq 0.6
%\eeq
%which results in $\langle \sigma v_{\rm rel} \rangle_{\rm tot}^{\rm bnd} \simeq 8.7 \times 10^{-24} {\rm cm}^3/{\rm sec}$.
%As an example, for $m_V = 1 \TeV$, one finds \cite{}
%\bea
%&f^{WW} = 0.19, \ f^{ZZ} = 0.18, \ f^{hh} = 0.22,&
%\nonumber \\
%&{\rm Br}^{WW} \simeq 0.5, \ {\rm Br}^{ZZ} \simeq 0.25, \ {\rm Br}^{hh} \simeq 0.25&
%\eea
%which results in $\langle \sigma v_{\rm rel} \rangle_{\rm tot}^{\rm bnd} \simeq 6.1 \times 10^{-23} {\rm cm}^3/{\rm sec}$.
The total annihilation cross section can be expressed as
\beq
\langle \sigma v_{\rm rel} \rangle_{\rm tot} = \langle \sigma v_{\rm rel} \rangle_0 \mathcal{S}
\eeq
where $\langle \sigma v_{\rm rel} \rangle_0$ is the perturbative annihilation cross section.
Then, requiring $\langle \sigma v_{\rm rel} \rangle_{\rm tot} < \langle \sigma v_{\rm rel} \rangle_{\rm tot}^{\rm cmb}$, we find
\bea
\frac{\langle \sigma v_{\rm rel} \rangle_0}{\langle \sigma v_{\rm rel} \rangle_{26}}
&<& \frac{20/\mathcal{S}}{\sum_{i={\rm channels}} f_{\rm eff, sys}^i {\rm Br}^i} \l( \frac{m_V}{1 \TeV} \r)
\nonumber \\
&\simeq& \frac{(20/\pi) \times 10^{-6}}{\sum_{i={\rm channels}} f_{\rm eff, sys}^i {\rm Br}^i} \l( \frac{v_{\rm cmb}}{10^{-10}} \r) \l( \frac{10^{-5}}{\alpha_X} \r) \l( \frac{m_V}{100 \GeV} \r) 
\eea
where $\langle \sigma v_{\rm rel} \rangle_{26} \equiv 6 \times 10^{-26} {\rm cm}^3/{\rm sec}$, and we used $\mathcal{S} \approx \pi \alpha_X / v_{\rm cmb}$ in the second line.%\eqs{alphaX-bound}{vdm-cmb-2} in the second line.
%---------------------
\begin{figure}
\centering
\includegraphics[width=0.45\textwidth]{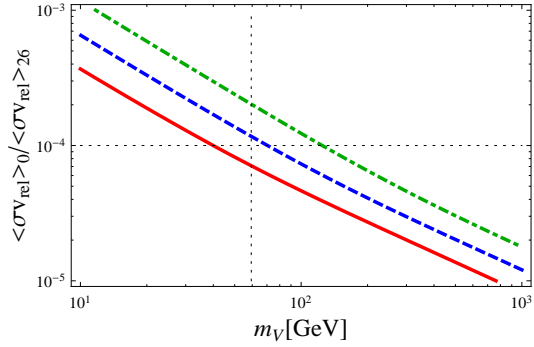}
\caption{\label{fig:sv0bnd-cmb} 
Upper bound of $\langle \sigma v_{\rm rel} \rangle_0/\langle \sigma v_{\rm rel} \rangle_{26}$ around the epoch of CMB decoupling. 
Solid red, dashed blue and dot-dashed green correspond to $\alpha_X$ saturating the upper-bound $a \times \l( \sigma_T / m_V \r)^{\rm max}$ with $a =1, 10^{-1}, 10^{-2}$, respectively.
We used $\sum_{i={\rm channels}} f_{\rm eff, sys}^i {\rm Br}^i=0.1$.}
\end{figure}
%---------------------
%
%Now we can use \eq{alphaX-bound} for \eq{Tprime-kd} to estimate the photon temperature when DM get through eventual kinetic decoupling.
%In this case, one finds 
%\beq
%T'_{\rm kd} \gtrsim 19 \MeV, \ T_\gamma' \gtrsim 10 \MeV
%\eeq
%where we used $g_{*S}(T_{\rm kd}) = 75.75$ and $g_{*S}(T_{\rm kd}')=10.75$.
%This photon temperature leads
%\beq \label{vdm-cmb-2}
%v(t_{\rm cmb}) \lesssim 3.0 \times 10^{-10} \l( \frac{19 \MeV}{T_{\rm kd}'} \r)^{1/2} \l( \frac{m_h/2}{m_V} \r)^{1/2}
%\eeq
%
Fig.~\ref{fig:sv0bnd-cmb} shows that $\langle \sigma v_{\rm rel} \rangle_0$ around the epoch of CMB decoupling should be smaller than $\langle \sigma v_{\rm rel} \rangle_{26}$ by several orders of magnitude.

A pair of $V^+$-$V^-$ can annihilate via the $s$-channel $\phi$ exchanges and its mixing with the SM Higgs boson $h$, and $t$-channel $V^\pm$ exchanges.
Hence the total annihilation cross section without nonperturbative enhancement effect taken into account is given by
\bea
%\langle \sigma v_{\rm rel} \rangle_0 
%&=& \langle \sigma v_{\rm rel} \rangle_{VV \to {\rm SM}}^{s} + \langle \sigma v_{\rm rel} \rangle_{VV \to \gamma_h \gamma_h}^t 
%\nonumber \\
%&& + \langle \sigma v_{\rm rel} \rangle_{VV \to H_1 H_1}^t + \langle \sigma v_{\rm rel} \rangle_{VV \to H_2 H_2}^t + \langle \sigma v_{\rm rel} \rangle_{VV \to H_1 H_2}^t
\l( \sigma v_{\rm rel} \r)_0 
&=& \l( \sigma v_{\rm rel} \r)_{VV \to {\rm SM}}^{s} + \l( \sigma v_{\rm rel} \r)_{VV \to \gamma_h \gamma_h}^t 
\nonumber \\
&& + \l( \sigma v_{\rm rel} \r)_{VV \to H_1 H_1}^t + \l( \sigma v_{\rm rel} \r)_{VV \to H_2 H_2}^t + \l( \sigma v_{\rm rel} \r)_{VV \to H_1 H_2}^t
\eea
where
\bea
\l( \sigma v_{\rm rel} \r)_{VV \to {\rm SM}}^{s} 
&=& \frac{1}{2 s} m_V^2 \alpha_X \sin^2 (2 \alpha) \l| \sum_{i=1,2} \frac{(-1)^{i+1}}{s - m_i^2 + i\Gamma_i m_i} \r|^2 \l[ 12 - 4 \l( \frac{s}{m_V^2} \r) + \l( \frac{s}{m_V^2} \r)^2 \r]
\nonumber \\
&& \l\{ \frac{1}{144} \l( 1 - \frac{4 m_W^2}{s} \r)^{1/2} m_W^2 g^2 \l[ 12 - 4 \l( \frac{s}{m_W^2} \r) + \l( \frac{s}{m_W^2} \r)^2 \r] \r.
\nonumber \\
&& \l. + \frac{1}{2 \times 144} \l( 1 - \frac{4 m_Z^2}{s} \r)^{1/2} m_Z^2 \l( \frac{g}{\cos \theta_W} \r)^2 \l[ 12 - 4 \l( \frac{s}{m_Z^2} \r) + \l( \frac{s}{m_Z^2} \r)^2 \r] \r.
\nonumber \\
&& \l. + \frac{1}{18} \sum_f N_{{\rm c},f}\l( 1 - \frac{4 m_f^2}{s} \r)^{3/2} \l( \frac{m_f}{2 m_W} \r)^2 g^2 s \r\}
\\
\l( \sigma v_{\rm rel} \r)_{VV \to {\rm H_i H_j}}^{s}
&=& 
\frac{\mathcal{S}_{ij} \alpha_X m_V^2}{36 \pi s} \l| \frac{s_\alpha \lambda_{1ij}}{s - m_1^2 + i \Gamma_1 m_1} + \frac{c_\alpha \lambda_{2ij}}{s - m_2^2 + i \Gamma_2 m_2} \r|^2
\nonumber \\
&& \l[ 12 - 4 \l( \frac{s}{m_V^2} \r) + \l( \frac{s}{m_V^2} \r)^2 \r] \l[ 1 + \l( \frac{m_i^2 - m_j^2}{s} \r)^2 - 2 \l( \frac{m_i^2+m_j^2}{s} \r) \r]^{1/2}
\\
\l( \sigma v_{\rm rel} \r)_{VV \to \gamma_h \gamma_h}^t 
&=& \frac{\pi \alpha_X^2}{9 m_V^2 \beta} \l[ 44 \beta -18 \beta^3 +6 \beta^5 -3 (1-\beta^2)^2 (1+\beta^2) \log\frac{1+\beta}{1-\beta} \r]
\\
\l( \sigma v_{\rm rel} \r)_{VV \to H_1 H_1}^t 
&=& \frac{\pi}{s} \alpha_X^2 \sin^4 \alpha \l( 1 - \frac{4 m_1^2}{s} \r)^{1/2} \l[ 12 - 4 \l( \frac{s}{m_V^2} \r) + \l( \frac{s}{m_V^2} \r)^2 \r]
\nonumber \\
\l( \sigma v_{\rm rel} \r)_{VV \to H_2 H_2}^t 
&=& \frac{\pi}{s} \alpha_X^2 \cos^4 \alpha \l( 1 - \frac{4 m_2^2}{s} \r)^{1/2} \l[ 12 - 4 \l( \frac{s}{m_V^2} \r) + \l( \frac{s}{m_V^2} \r)^2 \r]
\nonumber \\
\l( \sigma v_{\rm rel} \r)_{VV \to H_1 H_2}^t 
&=& \frac{\pi}{2s} \alpha_X^2 \sin^2(2 \alpha) \l[ 1 - \frac{\l(m_1-m_2\r)^2}{s} \r]^{1/2} \l[ 1 - \frac{\l(m_1+m_2\r)^2}{s} \r]^{1/2} 
\nonumber \\
&& \times \l[ 12 - 4 \l( \frac{s}{m_V^2} \r) + \l( \frac{s}{m_V^2} \r)^2 \r]
\eea
with $\alpha$ and $N_{{\rm c},f}$ being respectively the mixing angle and the color factor of the SM fermion $f$, $\mathcal{S}_{ij} = (1/2,1)$ for $(i=j, i\neq j)$,  
\bea
\lambda_{111} 
&=& 6 \lambda_\phi v_\phi s_\alpha^3 + 3 \lambda_{\phi H} v_H c_\alpha s_\alpha^2 + 3 \lambda_{\phi H} v_\phi c_\alpha^2 s_\alpha + 6 \lambda_H v_H c_\alpha^3
\\
\lambda_{122}
&=& \lambda_{\phi H} v_H c_\alpha^3 - 2 \l( -3 \lambda_H + \lambda_{\phi H} \r) v_H c_\alpha s_\alpha^2 - 2 \l( -3 \lambda_\phi + \lambda_{\phi H} \r) v_\phi c_\alpha^2 s_\alpha + \lambda_{\phi H} v_\phi s_\alpha^3
\\
\lambda_{211}
&=& \lambda_{\phi H} v_\phi c_\alpha^3 - 2 \l( -3 \lambda_\phi + \lambda_{\phi H} \r) v_\phi c_\alpha s_\alpha^2 + 2 \l( -3 \lambda_H + \lambda_{\phi H} \r) v_H c_\alpha^2 s_\alpha - \lambda_{\phi H} v_H s_\alpha^3
\\
\lambda_{222} 
&=& 6 \lambda_\phi v_\phi c_\alpha^3 - 3 \lambda_{\phi H} v_H c_\alpha^2 s_\alpha + 3 \lambda_{\phi H} v_\phi c_\alpha s_\alpha^2 - 6 \lambda_H v_H s_\alpha^3
\eea
and $\beta =(1-4 m_V^2/s)^{1/2}$.
We notice that, for $\alpha_X \lesssim \alpha_X^{\rm max}$, the total annihilation cross 
section is generically much smaller than $\langle \sigma v_{\rm rel} \rangle_{26}$.
However, as shown in Fig.~\ref{fig:svth-vs-mV-Rhiggs}, $\langle \sigma v_{\rm rel} \rangle_{\rm tot}$ can reach $\langle \sigma v_{\rm rel} \rangle_{26}$ around the 
$s$-channel resonance region. 
%---------------------
\begin{figure}
\centering
\includegraphics[width=0.45\textwidth]{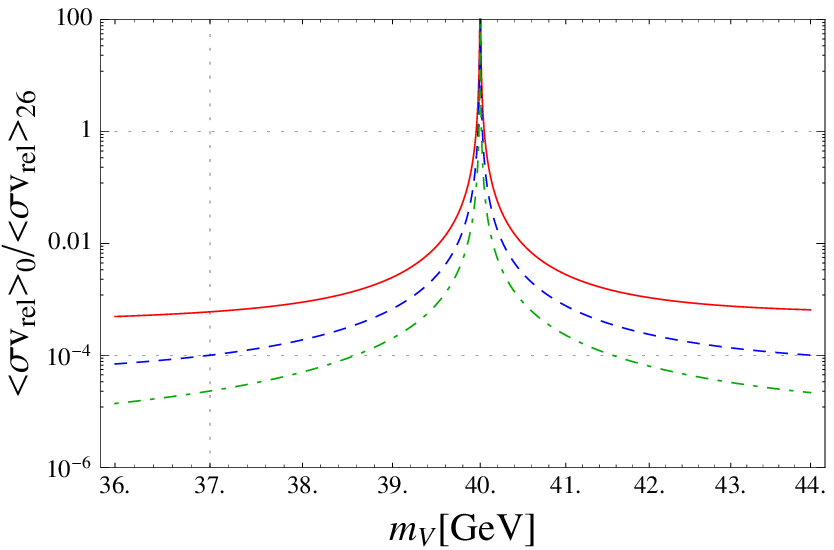}
\includegraphics[width=0.45\textwidth]{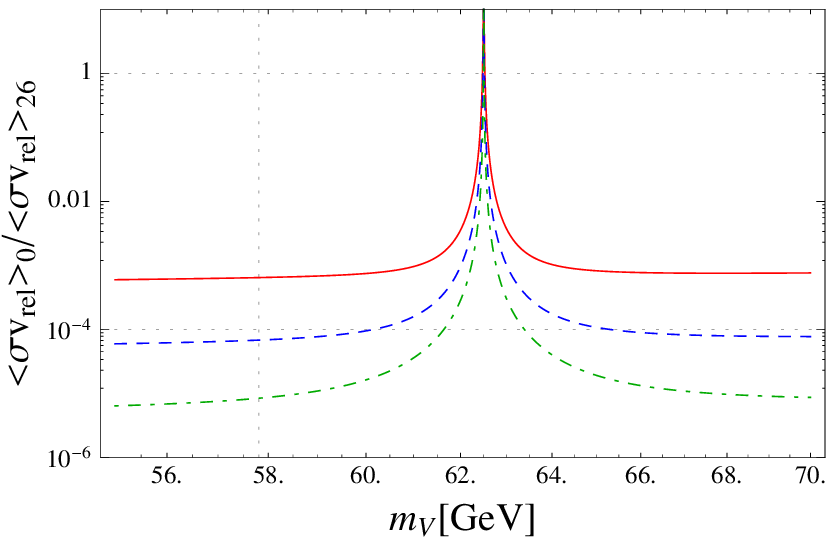}
\caption{\label{fig:svth-vs-mV-Rhiggs} 
Velocity-averaged annihilation cross section at the CMB decoupling epoch, normalized by the canonical value for thermal relic (i.e., $\langle \sigma v_{\rm rel} \rangle_{\rm fz} = 6 \times 10^{-26} {\rm cm}^3/{\rm sec}$).
Left: $m_\phi = 80 \GeV$ and $\alpha=0.25$.
Right: $m_\phi = 200 \GeV$ and $\alpha=0.1$.
Colored lines corresponds to $a = 1,10^{-1},10^{-2}$ from top to bottom with $a$ defined in Fig.~\ref{fig:sv0bnd-cmb}. 
Dashed vertical line corresponds to $\sqrt{s}/2 - m_V \simeq (3/2)T_{\rm fz}$ with $T_{\rm fz} = m_V/20$ for $\sqrt{s} = m_\phi, m_h$ in the left and right panel, respectively.}
\end{figure}
%---------------------
Hence, for the present relic density of VDM to be consistent with observation, the energy of VDM had to be in the resonance band at freeze-out.
It turned out that for $\sqrt{s} \approx m_\phi \gtrsim 150 \GeV$ the resonance band is not narrow enough to satisfy both of CMB constraint and relic density requirement simultaneously.
On the other hand, if $\sqrt{s} \approx m_\phi \lesssim 80 \GeV$ (or $\approx m_h$), 
and the dark gauge coupling $\alpha_X$ satisfies the following condition,  %if 
\beq
\alpha_X \lesssim \alpha_X^{\rm CMB} \equiv \alpha_X^{\rm max} / \sqrt{10} , 
\eeq
the resonance is quite sharp so that we can obtain a right amount of VDM relic density while 
satisfying the CMB constraint.
%Quantitatively, CMB constraint can be satisfied for $m_V \approx m_h/2 - (3/2)T_{\rm fz} - \Delta$ with 
%\beq
%g_X < g_X^{\rm max}, \ \alpha \lesssim 0.2, \ \Delta \sim 10^{-4}
%\eeq
%or for $m_V \approx m_\phi/2 - (3/2)T_{\rm fz} - \Delta$ with 
%\beq
%g_X \lesssim g_X^{\rm max}/2, \ \alpha \sim \alpha_{\rm SM-bnd}, \ \Delta \sim 10^{-4} - 10^{-3}
%\eeq
%In order to solve the small scale puzzles, we take $g_X \gtrsim g_X^{\rm max}/3$, and find that still $\Delta \sim 10^{-4}$ is enough for a right amount of relic density while CMB constraint is satisfied.

%-------------------------------------------------------------------------------------------------------------
%-------------------------------------------------------------------------------------------------------------
%\subsection
\subsection{Relic Densities of monopoles and VDMs} \label{sec:relics}

\subsubsection{VDMs}
The massive vector bosons $V^\pm$ make good CDM of the universe, and they are thermalized mainly by the Higgs portal interaction, $\lambda_{\phi H}$ term in \eq{Lag}. 
For $\alpha_X \lesssim \alpha_X^{\rm CMB}$ and small mixing angle $\alpha$, 
the annihilation cross section is typically much smaller than the canonical value for 
a right  amount of relic density at present except for the  $s$-channel resonance.  
On the $s$-channel resonance, for $|m_1 - m_2| \gg {\rm Max}[\Gamma_1, \Gamma_2]$ 
with $\Gamma_{1,2}$ being the total decay width of $H_{1,2}$, the present relic density 
can be approximated to \cite{Gondolo:1990dk} %formulae in  \cite{Gondolo:1990dk}.
\bea \label{eq:YV-res}
Y_{V,0}^{\rm res}
%&\approx& C_V \frac{m_V}{M_{\rm P}} \frac{m_{\rm R}}{\Gamma_{\rm R}} \frac{\epsilon_{\rm R}^{1/2}}{{\rm B}_i {\rm B}_f} \frac{\Theta(\epsilon_{\rm R})}{{\rm erfc}(\sqrt{x_{\rm fz} \epsilon_{\rm R}})}
%\nonumber \\
&\approx& C_V \frac{m_V/M_{\rm P}}{(3/4) f_{\rm R}(\alpha) \alpha_X} \frac{\Gamma_{\rm R}(\alpha)}{\Gamma_{\rm R}^{\rm SM}(\alpha)} \frac{\Theta(\epsilon_{\rm R})}{{\rm erfc}(\sqrt{x_{\rm fz} \epsilon_{\rm R}})}
%\nonumber \\
%&\gtrsim& 9.4 \times 10^{-15} \GeV \l( \frac{1 \TeV}{m_V} \r)^{1/2} \frac{\Gamma_{\rm R}}{\Gamma_{\rm R}^{\rm SM}} \frac{\Theta(\epsilon_{\rm R})}{{\rm erfc}(\sqrt{x_{\rm fz} \epsilon_{\rm R}})}
\eea
%where $Y_{V,0}^{\rm res}$ is the yield of VDM at present in the resonance band, $C_V \equiv \frac{27}{32 \pi^2} \sqrt{\frac{5}{2 g_*(T_{\rm fz})}}$ with $g_*(T_{\rm fz})\sim100$ being the relativistic degrees of freedom at freeze-out, $m_{\rm R}$ and $\Gamma_{\rm R}$ are respectively the mass and total decay with of resonance, $\epsilon_{\rm R} \equiv \l( 1 - 4 m_V^2/m_{\rm R}^2 \r) m_{\rm R}^2 / 4 m_V^2$, ${\rm B}_{i,f}$ is the decay branching fraction of the resonance to the initial/final states, $x_{\rm fz} \equiv m_V/T_{\rm fz} \sim 25$, $\Theta(x)$ is the Heaviside step-function, and %in the second line we used 
%
where $Y_{V,0}^{\rm res}$ is the yield of VDM at present in the resonance band, 
$C_V \equiv \frac{27}{32 \pi^2} \sqrt{\frac{5}{2 g_*(T_{\rm fz})}}$ with 
$g_*(T_{\rm fz})\sim100$ being the relativistic degrees of freedom at freeze-out, 
$m_{\rm R}$ and $\Gamma_{\rm R}$ are respectively the mass and total decay width of resonance, $\Theta(x)$ is the Heaviside step-function, 
$\epsilon_{\rm R} \equiv \l( 1 - 4 m_V^2/m_{\rm R}^2 \r) m_{\rm R}^2 / 4 m_V^2$, 
$x_{\rm fz} \equiv m_V/T_{\rm fz} \sim 25$, and the decay rates of a resonance to DM and 
SM particles are respectively given by
\bea
\Gamma_{\rm R}^{\rm DM}(\alpha) &\approx& f_{\rm R}(\alpha) \frac{3}{4} \alpha_X m_{\rm R} \epsilon_{\rm R}^{1/2}
\\
\Gamma_{\rm R}^{\rm SM}(\alpha) &=& \l( 1 - f_{\rm R}(\alpha) \r) \Gamma^{\rm SM}(m_{\rm R})
\eea
where $f_{\rm R}(\alpha) = (\sin^2 \alpha, \cos^2 \alpha)$ for ${\rm R} = (1,2)$, and $\Gamma^{\rm SM}(m_{\rm R})$ is the decay rate of SM Higgs in case $m_h$ is replaced to $m_{\rm R}$.
As shown in the left panel of Fig.~\ref{fig:alpha-vs-tuning-Rhiggs}, defining the level of 
tuning for resonance as $\Delta m_V/m_V \equiv \l(\sqrt{s} - m_{\rm R} \r)/\l(2 m_V \r)$, 
we find that a tuning smaller than about $\mathcal{O}(1)\%$ is necessary to obtain a right amount of relic density, 
if $m_V \approx m_h/2$ and $\alpha_X \lesssim \alpha_X^{\rm CMB}$.
In case of $m_V \approx m_\phi$, the fine tuning parameter is approximately 
given by $\Delta m_V/m_V = \mathcal{O}(1-10)$\% for 
$\alpha_X = \l(0.1-1 \r) \times \alpha_X^{\rm max}$. Note that it does not depend on 
the mixing angle $\alpha$,  since $\alpha$-dependence  is nearly 
cancelled out in \eq{eq:YV-res}.
%---------------------
\begin{figure}
\centering
\includegraphics[width=0.45\textwidth]{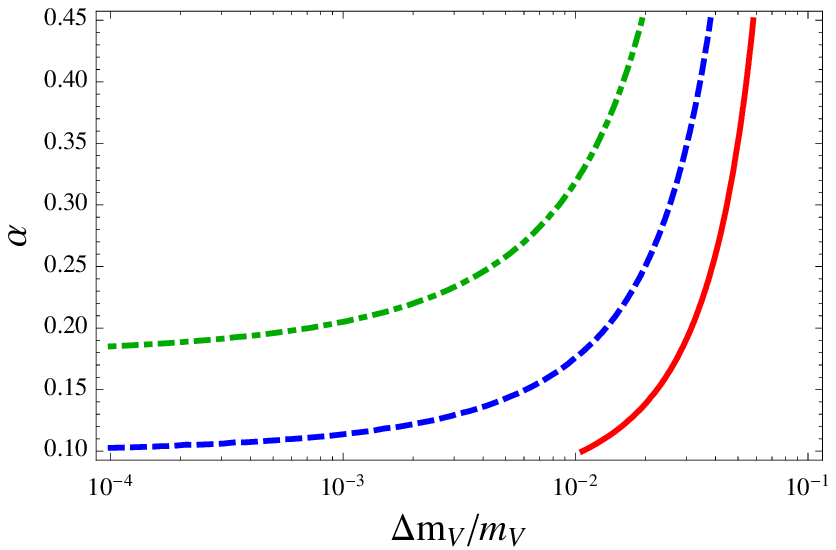}
\includegraphics[width=0.45\textwidth]{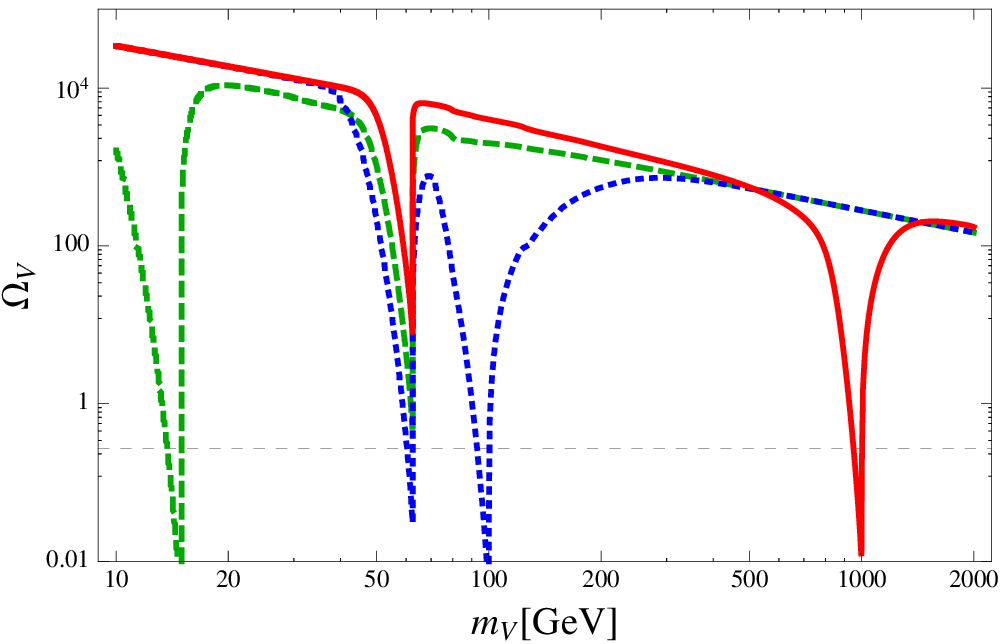}
\caption{\label{fig:alpha-vs-tuning-Rhiggs} 
Left: Contours of $\Omega_{\rm VDM} = \Omega_{\rm CDM}^{\rm obs}$ as a function of $\alpha$ and $\Delta m_V/m_V$ for $m_V \approx m_h/2 - (3/2) T_{\rm fz}$ and $a = 1,10^{-1},10^{-2}$ from right to left with $a$ defined in Fig.~\ref{fig:sv0bnd-cmb}.
Right: $\Omega_{\rm VDM}$ as a function of $m_V$ for $m_1=125 \GeV$ and $m_2 = 30 \GeV, 0.2 \TeV, 2 \TeV$ (dashed-green, dotted-blue, solid-red).
$g_X$ was chosen to be the half of the upper-bound in \eq{alphaX-bound}, and $\alpha = 0.3 m_1 m_2/(m_2^2-m_1^2)$ was used.
}
\end{figure}
%---------------------
The right panel of Fig.~\ref{fig:alpha-vs-tuning-Rhiggs} is the thermal relic density of hidden sector VDM, obtained by micromegas ~\cite{Belanger:2008sj}, as functions of $m_V$ for $m_2 = 0.03, 0.2, 2 \TeV$.
%
%%---------------------
%\begin{figure}
%\centering
%\includegraphics[width=0.45\textwidth]{figs/OmegaV-vs-mV.eps}
%\caption{\label{fig:OmegaV-vs-mV} The relic abundance of VDM as a function of $m_V$ for $m_1=125 \GeV$ and $m_2 = 30 \GeV, 0.2 \TeV, 2 \TeV$ (dashed-green, dotted-blue, solid-red).
%$g_X$ was chosen to be the half of the upper-bound in \eq{alphaX-bound}, and $\alpha = 0.3 m_1 m_2/(m_2^2-m_1^2)$ was used.
%Gray region may be excluded since vacuum stability of Higgs potential requires too large mixing.
%}
%\end{figure}
%%---------------------
%
%{\color{red} Quantitatively, CMB constraint can be satisfied for $m_V \approx m_h/2 - (3/2)T_{\rm fz} - \Delta$ with 
%\beq
%g_X < g_X^{\rm max}, \ \alpha \lesssim 0.2, \ \Delta \sim 10^{-4}
%\eeq
%or for $m_V \approx m_\phi/2 - (3/2)T_{\rm fz} - \Delta$ with 
%\beq
%g_X \lesssim g_X^{\rm max}/2, \ \alpha \sim \alpha_{\rm SM-bnd}, \ \Delta \sim 10^{-4} - 10^{-3}
%\eeq
%In order to solve the small scale puzzles, we take $g_X \gtrsim g_X^{\rm max}/3$, and find that still $\Delta \sim 10^{-4}$ is enough for a right amount of relic density while CMB constraint is satisfied.
%}

\subsubsection{Monopoles}
The $SU(2)$ symmetry is broken down to $U(1)_X$ as the phase transition takes place at temperature $T_{\rm c} \simeq \sqrt{\l| m_{\phi \phi}(\phi = 0)^2 \r|} / \sqrt{(5/12) \lambda_\phi + g_X^2/2}$ with $m_{\phi \phi}^2(\phi = 0) = - \lambda_\phi v_\phi^2$ being the zero temperature mass of $\phi$ at the origin.
The strength of the phase transition characterized by $\phi(T_{\rm c})/T_{\rm c}$ is \cite{Dine:1992vs}
\beq
\frac{\phi(T_{\rm c})}{T_{\rm c}} \approx \frac{2}{\lambda_\phi} \l( \frac{m_V^3}{3 \pi v_\phi^3} \r) = \frac{2 g_X^3}{3 \pi \lambda_\phi} 
\eeq
Hence, for $g_X$ satisfying \eq{alphaX-bound} with $m_V \lesssim \mathcal{O}(1) \TeV$ and $m_\phi \gtrsim \mathcal{O}(10) \GeV$, the phase transition is nearly of the second-order type (i.e. $\phi(T_{\rm c}) / T_{\rm c} \ll 1)$.
In this case, based on the Kibble-Zurek mechanism \cite{Kibble:1976sj,Zurek:1985qw}, the initial abundance of monopole at its formation is expected to be \cite{Murayama:2009nj}
\beq \label{Ym-ini}
Y_i \approx \frac{ \l( \sqrt{\lambda_\phi/2}\r)^3}{C_S} \l[ \frac{1}{\sqrt{\lambda_\phi/2}} C_0^{1/2} \frac{m_{\rm M}}{g_{\rm M} M_{\rm P}} \r]^{3 \nu/(1 + \mu)}
\eeq
where $Y_i \equiv n_i/s$ with $n_i$ and $s$ being respectively the number density of monopoles and entropy density, $C_S \equiv \l( \frac{2 \pi^2}{45} g_{*S} \r)$, $C_0 \equiv \l( \frac{\pi^2}{90} g_* \r)$ with $g_{*S}$ and $g_*$ being respectively the relativistic degrees of freedom associated with entropy and energy densities, and $g_{\rm M} \equiv 4 \pi/g_X$ is the magnetic charge of monopoles.
For a negligible Hubble expansion rate, the classical values of the critical exponents are $\nu = \mu = 1/2$, but quantum corrections increase them to $\nu = \mu = 0.7$ \cite{Murayama:2009nj}.
%Hence, for $\lambda_\phi \lesssim \mathcal{O}(0.1)$ and $m_V = \mathcal{O}(10-10^3) \GeV$ with \eq{alphaX-bound} satisfied, $Y_i$ is too small to provide a sizable contribution to DM relic density at present, as shown in Fig.~\ref{fig:OmegaM-vs-mV}.
%

The initial abundance of monopoles may be reduced further by monopole-antimonopole annihilation resulting from formation of a magnetic Coulomb bound state and its subsequent cascade decays into dark photons \cite{Preskill:1979zi}.
However in our scenario the annihilation caused by such collisions are not efficient 
enough to reduce monopole density further than that in \eq{Ym-ini}.
This is because in the thermal bath only $V^\pm$ which have dark charge interact 
directly with monopoles.
In order to reduce monopole density significantly, $V^\pm$ should be light enough.
This means that for a given $v_\phi$, $g_X$ should be small enough, but this results 
in very heavy monopoles and the effect of $g_X$ to the fractional energy density of 
monopoles is cancelled out.
In addition, $v_\phi$ is subject to the constraint \eq{alphaX-bound}, and $g_X$ should 
not be too small to avoid over-production of dark matter.  
Hence, $v_\phi$ can not be small and, as the result, the energy contribution of monopole 
can not be reduced much.
One may expect that the mixing between dark and visible sector Higges may results 
in a sizable reduction of monopole density due to the interactions to SM particles.
However note that the mixing angle $\alpha$ is constrained to be less than about $0.45$ 
for 95\% CL \cite{Chpoi:2013wga}.  So, its effect is at most comparable to the case of
$V^\pm$ for $m_V \lesssim \mathcal{O}(1) \TeV$.

%---------------------
\begin{figure}
\centering
\includegraphics[width=0.45\textwidth]{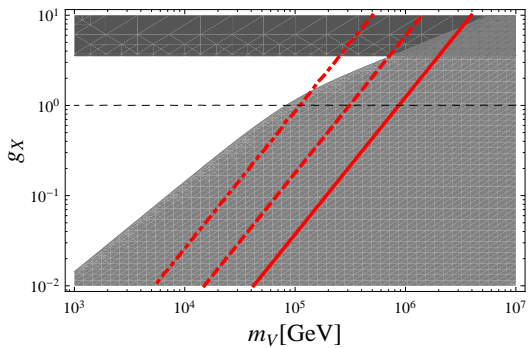}
\caption{\label{fig:OmegaM-in-gX-mV} Contours of the relic abundance of monopoles, corresponding to $\Omega_M/\Omega_{\rm CDM} = 10^{-2}, 0.1, 1$ for dot-dashed, dashed and solid lines, respectively.
Gray and dark-gray regions are excluded due to too much VDM and perturbativity of gauge coupling, respectively.
\eq{eq:YV-res} was used for the VDM relic density with $m_\phi = 2 m_V$ and $\Delta m_V / m_V = 10^{-4}$.
}
\end{figure}
%---------------------
Fig.~\ref{fig:OmegaM-in-gX-mV} shows contours of thermal relic density of hidden sector monopoles as a function of $g_X$ and $m_V$, which was obtained from \eq{Ym-ini}.
We notice that the monopole abundance can be about 10\% of the observed dark matter 
relic density at best in the price of $10^{-2}$\% tuning of $m_V$ at $\mathcal{O}(1) \PeV$ 
scale, even though we take $\alpha_X \sim \mathcal{O}(1)$ for which  perturbative 
description  of our model may be not valid any more.
Hence VDM should be the main component of the present dark matter relic density, 
and in this case the mass of VDM is constrained to be close to or smaller than $m_h/2$ 
as discussed in the previous subsections.
The relic abundance of monopoles turns out to be $\Omega_{\rm M} \sim \mathcal{O}(10^{-6}-10^{-5}) \Omega_{\rm CDM}$ for such a light VDM, which is totally negligible.
Still the existence of hidden sector monopole is very crucial to guarantee the hidden sector
VDM $V_\mu^\pm$ to be absolutely stable due to the unbroken $U(1)_X$ subgroup of the original $SU(2)_X$.  
\subsection{Direct Detection}
\subsubsection{VDMs} 
As in the Abelian VDM case, the direct detection cross section of VDM-nucleon scattering occurs through the $t$-channel exchange of $H_1$ and $H_2$.
Due to the generic destructive interference between two scalar bosons, the strong bounds 
from the CDMS and XENON100 can be significantly relaxed if $m_1 \sim m_2$, 
as shown in Ref.~\cite{vector_dm}. 

The spin-independent elastic cross section $\sigma_p$ of the VDM $V^\pm$ 
scattering off the proton is obtained as \cite{vector_dm}
\begin{equation}
\sigma_p = \frac{4 \mu_V^2}{\pi} \left(\frac{g_X s_\alpha c_\alpha m_p}{2 v_H}\right)^2
\left(\frac{1}{m_1^2}-\frac{1}{m_2^2}\right)^2 f_p^2,
\end{equation}
where $\mu_V =m_V m_p/(m_V + m_p)$, $m_p$ being the proton mass, and $f_p = \sum_{q=u,d,s} f_q^p + 2/9(1-\sum_{q=u,d,s} f_q^p) \approx 0.468$~\cite{Belanger:2008sj}. 
%---------------------
\begin{figure}
\centering
\includegraphics[width=0.45\textwidth]{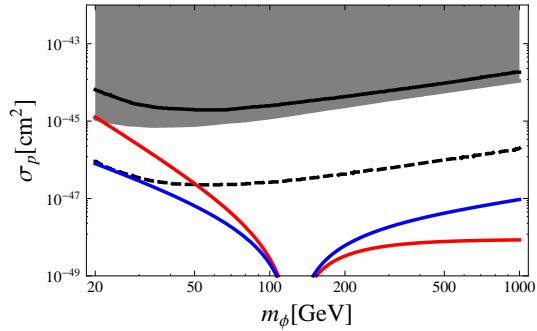}
\caption{\label{fig:sp-vs-mphi} Spin-indenpendent direct detection cross section of VDM 
(red and blue lines) at each resonance as a function of $m_\phi$ in the small mixing limit 
($\alpha = 0.1$) with $\alpha_X = \alpha_X^{\rm max}$.
Red lines: $m_V \approx m_h/2$ and we chose $\lambda_H = 0.140$ for the vacuum 
stability of SM Higgs potential.
Blue lines: $m_V \approx m_\phi/2$. 
Light gray region is excluded by LUX  experiment.
%Gray region may be excluded since vacuum stability of Higgs potential requires too large mixing.
The solid and the dashed black line is the XENON100 bound and the projected bound of 
XENON1T experiment.
Regions below red and blue lines in each case of resonance are consistent with \eq{alphaX-bound}.
}
\end{figure}
%---------------------
As an example, Fig.~\ref{fig:sp-vs-mphi} depicts the spin-independent direct detection 
cross section $\sigma_p$ for each resonance with $\alpha = 0.1$ (which can be valid 
up to $m_2 \simeq 1 \TeV$) and $g_X$ saturating the bound in \eq{alphaX-bound}.
As shown in the figure, all region of $m_\phi \gtrsim 20 \GeV$ can satisfy XENON100 
and LUX bounds \cite{Akerib:2013tjd} which is the most strong constraint as of now.
In addition, XENON1T \cite{Aprile:2012zx} may probe $m_\phi \lesssim 50 \GeV$ for 
$\sqrt{s} \approx m_h$ in optimistic cases. 

\subsubsection{Monopoles}
The monopole in our scenario is neutral under SM gauge group, and it interacts with the 
SM particles only via Higgs mediation thanks to the Higgs portal interaction we newly 
introduced in this paper.
The interaction is similar to that of DM-nucleon scattering via Higgs mediation.
However monopoles may be regarded as a particle located at the classical 
field $\phi_c$ (the solution for the classical field equation describing a monopole 
configuration) vanishes. 
%of $\phi$.} , % $\langle \phi \rangle \approx 0$}.
Then, the monopole-nucleon scattering cross section may be given by 
\beq
\frac{d \sigma_p}{d \Omega} 
= \frac{\mu_{\rm M}^2}{ 4 \pi^2} \l[ \frac{\lambda_{\Phi H}}{8} \frac{m_p}{m_{\rm M}} \l| 
\frac{f_p}{t-m_h^2} \r| \r]^2
\eeq
where $\mu_{\rm M} = m_{\rm M} m_p / \l( m_{\rm M} + m_p \r)$ with $t$ is the Mandelstam variable. %$m_p$ being the proton mass, $t$ is the Mandelstam variable, and $f_p = \sum_{q=u,d,s} f_q^p + 2/9(1-\sum_{q=u,d,s} f_q^p) \approx 0.468$~\cite{Belanger:2008sj}. 
Note that, for the recoil energy $E_{\rm r}$ of target atom in a direct search experiment,
 $t = - 2 m_A E_{\rm r}$ with $m_A$ being the mass of target atom.
Hence, for $E_{\rm r} \lesssim m_h^2 / (2 m_A)$, we find
\bea
\sigma_p 
&\lesssim& \frac{\lambda_{\Phi H}^2}{64 \pi m_{\rm M}^2} \l( \frac{m_p}{m_h} \r)^4 f_p^2
\nonumber \\
&\simeq& \frac{3.4 \times 10^{-28}}{\GeV^2} \l( \frac{\lambda_{\Phi H}}{0.1} \r)^2 \l( \frac{10^7 \GeV}{m_{\rm M}} \r)^2
\eea
It is far below sensitivities of present or near-future direct search experiments. 

%-------------------------------------------------------------------------------------------------------------
\section{Dark Radiation} \label{sec:DR}
The massless dark photon associated with the unbroken dark $U(1)_X$ symmetry 
contributes to the extra relativistic degrees of freedom in the present universe.
Starting from thermal equilibrium at high temperature, it is decoupled from VDM at 
$T_\gamma \sim 10 \MeV$ for a maximally allowed $g_X$ in \eq{alphaX-bound} 
\cite{Feng:2009mn}.
However, VDM is decoupled from SM thermal bath at much higher temperature since the thermal equilibrium of VDM is maintained only by Higgs mediation.

For a relativistic particle in thermal bath, the thermal-averaged scattering cross section of dark matter to a SM fermion is found to be 
\beq
\langle \sigma v \rangle_f 
\simeq 2 \alpha_X s_\alpha^2 c_\alpha^2 \l| \frac{1}{m_1^2} - \frac{1}{m_2^2} \r|^2 \frac{m_f^2}{v_H^2} E_f^2
\eeq
where we assumed that the momentum transfer is negligible relative to $m_1$ and $E_f$ is the energy of the SM fermion.
The scattering rate of DM to SM particles is then given by
\beq
\Gamma_s = \sum_f n_f \langle \sigma v \rangle_f
\eeq
where $f$ represents a SM fermion, and $n_f$ is its number density.
The kinetic decoupling takes place as the scattering rate of DM to SM particles becomes smaller than the Hubble expansion rate.
Let us  take the scalar mixing angle $\alpha$ to be $\alpha = 0.1$ for simplicity.
%---------------------
\begin{figure}
\centering
\includegraphics[width=0.45\textwidth]{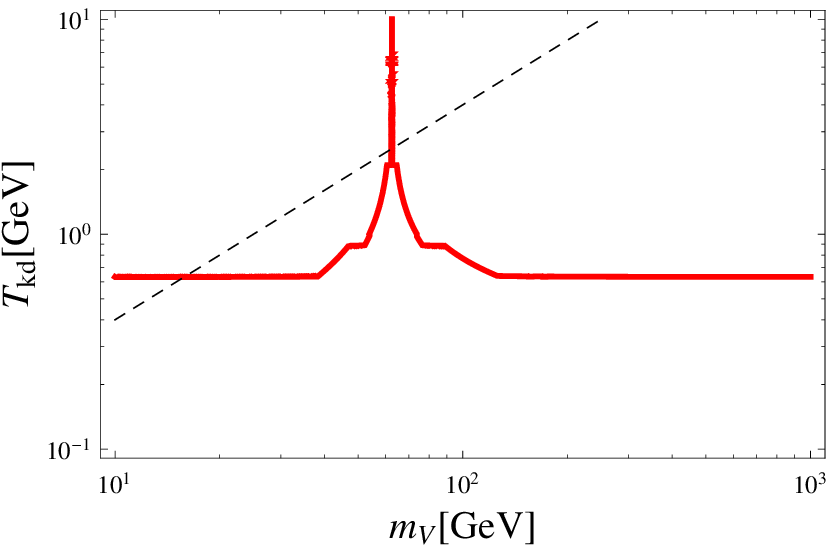}
\includegraphics[width=0.45\textwidth]{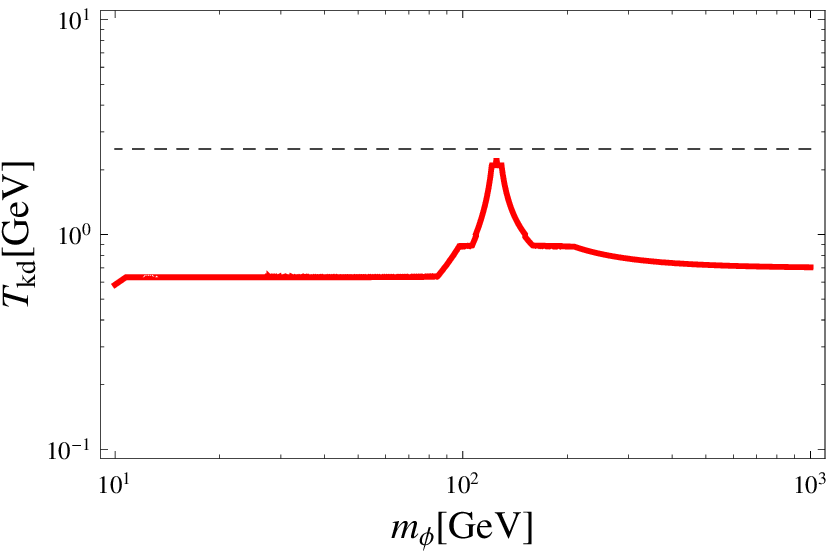}
\caption{\label{fig:Tkd-DR} 
Temperatures of photon at the kinetic decoupling of dark matter from SM thermal bath for $\alpha = 0.1$. 
Left: $\sqrt{s}=m_\phi$.
Right: $\sqrt{s}=m_h$.
We used $\alpha_X = \alpha_X^{\rm max}/\sqrt{10}$.
Dashed lines are $T_{\rm fz} = m_V/25$ for each resonance.
%Gray region may be excluded due to vacuum instability of SM higgs potential.
}
\end{figure}
%---------------------
Then, as shown in Fig.~\ref{fig:Tkd-DR}, $T_{\rm QCD} < T_{\rm kd} < 1 \GeV$ for most of region.
Even for $m_1 \sim m_2$, we find $T_{\rm kd} < T_{\rm fz} \sim 2 \GeV$ except the case of very high degeneracy. %, although the region may be likely to be excluded due to vacuum instability.
%When $\sqrt{s} \approx m_h$ (right panel), $T_\gamma^{\rm kd} < T_{\rm fz}$ for $m_\phi \lesssim 80 \GeV$, and $T_\gamma^{\rm kd}$ can reach the temperature of QCD-phase transition ($T_{\rm QCD} \approx 200 \MeV$) for $m_\phi \lesssim 20 \GeV$, but it does not go below since there are no light particles having large enough coupling to Higgs.
%For $m_\phi \gtrsim 80 \GeV$, $T_{\rm kd} > T_{\rm fz} \sim \mathcal{O}(1) \GeV$.
Note that $T_{\rm kd} > T_{\rm fz}$ for $m_V \sim 10 \GeV$.
In this case, VDM is in chemical equilibrium even after kinetic decoupling.
Since, as shown in Section~\ref{subsec:CMB-const}, VDM interacts strongly enough with dark photon to keep kinetic equilibrium, its kinetic energy at its production from SM thermal bath is redistributed among VDM and dark photons.
However, the kinetic energy is negligible relative to the energy of dark photon, hence temperature of dark photon is not changed.
%
%Additionally, there is no change of relativistic degrees of freedom for $T_{\rm QCD} < T < 1 \GeV$.
%Hence VDM can have equilibrium distribution, and $T_{\rm fz}$ is obtained in a usual manner.
%This implies that effectively $T_{\rm kd}$ becomes same as $T_{\rm fz}$. 
%
Therefore, we can say that the kinetic decoupling temperature is $T_{\rm QCD} < T_{\rm kd} \lesssim \mathcal{O}(1) \GeV$, and the contribution of dark photon to the present radiation energy density as the extra neutrino species can be $\Delta N_{\rm eff}^\nu \simeq 0.08 - 0.11$.
It is consistent with the recent result of Planck satellite mission \cite{Ade:2013zuv}, and can be probed at a Stage-IV CMB experiment at 2$\sigma$-level \cite{Abazajian:2013oma}.
Note that we could make a definite prediction to the amount of dark radiation from massless dark photon, based on an unbroken local dark gauge symmetry and thermal VDM with a Higgs portal interaction.
This is in sharp contrast to other models for extra dark radiations from, for example, axions or sterile neutrinos whose energy densities are usually adjusted by hand to match observations.

%{\color{blue} The mixing angle $\alpha$ might be restricted to be much smaller than $\mathcal{O}(0.1)$ in the future experiments, and $T_{\rm kd}$ may become quite high.
%However we still have $T_{\rm kd, eff} \leq T_{\rm fz}$ as far as VDM is in kinetic equilibrium with dark photon and the relativistic degrees of freedom is not changed for $T_{\rm fz} \leq T \leq T_{\rm kd}$.
%If $g_*(T_\gamma)$ is changed for $T < T_{\rm kd}$, there is a chance of energy transfer from SM thermal bath to dark photon via dark matter.
%The reason is as follows.
%The annihilation of VDM which is in kinetic equilibrium with dark photon freezes out at $T_{\gamma'} \sim m_V/25$. 
%However the production of VDM is closed later since $T > T_{\gamma'}$ as $g_*(T)$ is changed.
%Therefore, VDM is expected to be produced some amount after its annihilation freezes out.
%The energy of dark matter is redistributed with dark photon since they are still in kinetic equilibrium.
%As the result, the temperature of dark photon increases.
%A full quantitative analysis of this effect is out of the scope of this paper.
%}

%%%%%%%%%%%%%%%%%%%%%%%%%%%%%%%%%%%%%%%%%%%%%%%%---------------------------------------------------------------------------------------------------------------------
\section{Conclusions} \label{sec:conc}
%---------------------------------------------------------------------------------------------------------------------
In this paper, as a logically natural extension of SM to a dark sector in regard of local gauge 
principle and the stability of dark matter, we considered a $SU(2)$ dark gauge symmetry 
which is broken to $U(1)_X$ by a $SU(2)$-triplet dark Higgs field, which is the  
%The simplest realization of such a gauge group is the 
t'Hooft-Polyakov monopole model in the dark sector with Higgs portal.
% which consists of gauge fields and dark Higgs only.   
In this model, the dark sector consists of  monopoles and massive vector bosons,  
both of which are stable due to topology and unbroken $U(1)_X$ respectively, 
and massless dark photons.  
Although the constraint from CMB data on the dark matter annihilation cross section 
is quite stringent,  we showed that a right amount of thermal relic density can be obtained 
by resonant thermal freeze-out of massive VDM when the mass of VDM is 
close to or smaller than one for SM Higgs resonance, thanks to the Higgs portal 
which was newly introduced in this work.

The abundance of monopoles turns out to be negligible in this case, and their role
as CDM is not very important.  
However their existence is crucial for the VDM  to be
absolutely stable in the presence of higher dimensional nonrenormalizable operators,
due to the unbroken $U(1)_X$.
Present direct searches do not constrain VDM mass of $\mathcal{O}(10-10^3) \GeV$. 
But XENON1T experiment for example may probe VDM mass less than about $60 \GeV$.
The massless dark photon associated with the $U(1)_X$ contributes to the present 
radiation energy density, resulting in $\Delta N_{\rm eff}^\nu \sim 0.1$ as the extra 
relativistic neutrino species.   

Our hidden monopoles are quite rare and their scattering rate to nucleons looks too 
small to be detected at direct search experiments.
However, the self-interaction of monopoles, characterized by $g_{\rm M}\equiv 4 \pi/g_X$ is quite large for a small $g_X$, and it may cause monopoles captured at astrophysical 
object like sun.  In this case, the captured monopoles may be able to annihilate, and leave 
observable imprints.   We will take a look at this possibility in other place.

%
%\appendix
%\section{Some title}
%Please always give a title also for appendices.
%

\acknowledgments

We are grateful to Yuji Omura for collaboration at the initial stage, and to B. Holdom, 
Kimyeong Lee and E. Weinberg for useful discussion. 
This work was supported by NRF Research Grant 
2012R1A2A1A01006053, and by SRC program of NRF 
Grant No. 20120001176 funded by MEST through 
Korea Neutrino Research Center at Seoul National University.

%\paragraph{Note added.} This is also a good position for notes added
%after the paper has been written.

% The bibliography will probably be heavily edited during typesetting.
% We'll parse it and, using the arxiv number or the journal data, will
% query inspire, trying to verify the data (this will probalby spot
% eventual typos) and retrive the document DOI and eventual errata.
% We however suggest to always provide author, title and journal data:
% in short all the informations that clearly identify a document.

\end{document}